\pdfoutput=1

\documentclass[11pt,twoside,a4paper,cmspaper,final,collab]{cms-tdr}
\def\svnVersion{50968:51001}\def\cmsCernNo{2011-043}\def\cmsCernDate{\today}\def\cmsMessage{Submitted to Physical Review Letters}

\begin{document}\cmsNoteHeader{EWK-10-014}

\hyphenation{had-ron-i-za-tion}
\hyphenation{cal-or-i-me-ter}
\hyphenation{de-vices}
\RCS$Revision: 51001 $
\RCS$HeadURL: svn+ssh://alverson@svn.cern.ch/reps/tdr2/papers/EWK-10-014/trunk/EWK-10-014.tex $
\RCS$Id: EWK-10-014.tex 51001 2011-04-19 17:14:17Z alverson $
\newcommand\rs{\raisebox{1.0ex}[-1.0ex]}
\newcommand{\ra}{\ensuremath{\rightarrow}}
\newcommand{\znunu}{\ensuremath{{\text Z} \ra \nu\bar{\nu}}}
\newcommand{\zmumu}{\ensuremath{{\text Z} \ra \mu\mu}}
\newcommand{\wmunu}{\ensuremath{{\text W} \ra \mu\nu}}
\newcommand{\wtaunu}{\ensuremath{{\text W} \ra \tau\nu}}
\newcommand{\dphi}{\ensuremath{\Delta \phi}}
\newcommand{\dphijj}{\ensuremath{\Delta \phi_{ j1,j2}}}
\newcommand{\Pt}{\ensuremath{{p_{\text T}}\xspace}}
\newcommand{\pts}{\ensuremath{p_{\text T}{\text s}}\xspace}
\newcommand{\Et}{\ensuremath{{E_{\text T}}\xspace}}
\newcommand{\ptjf}{\ensuremath{p_{\rm T}^{ {\rm j}_1} }}
\newcommand{\ptjs}{\ensuremath{p_{\rm T}^{ {\rm j}_2} }}
\newcommand{\ptjt}{\ensuremath{p_{\rm T}^{ {\rm j}_3} }}
\newcommand{\etajf}{\ensuremath{\eta^{ {\rm j}_1} }}
\newcommand{\etajs}{\ensuremath{\eta^{ {\rm j}_2} }}
\newcommand{\etajt}{\ensuremath{\eta^{ {\rm j}_3} }}
\newcommand{\ttj}{\ensuremath{\rm{t}\bar{\rm{t}} + jets}\xspace}
\newcommand{\wj}{\ensuremath{\rm W + jets}\xspace}
\newcommand{\zj}{\ensuremath{\rm Z + jets}\xspace}
\newcommand{\etaabs}{\ensuremath{|\eta|}}
\newcommand{\mjj}{\ensuremath{M_{\text{inv}}^{j1,j2}}}
\newcommand{\chiznew}{\ensuremath{\chi^{0}}\xspace}
\newcommand{\chipnew}{\ensuremath{\chi^{+}}\xspace}
\newcommand{\sQuanew}{\ensuremath{\tilde{\rm q}}\xspace}
\newcommand{\sGlunew}{\ensuremath{\tilde{\rm g}}\xspace}
\newcommand{\ttNew}{\ensuremath{\rm{t}\bar{\rm{t}}}\xspace}
\newcommand{\tev}{\TeV}

\newcommand{\combIso}{Iso_{\textrm{comb.}}}
\renewcommand{\arraystretch}{1.2}
\newcommand{\bigNum}[2]{#1 \, \times \, 10 \, ^{#2}}

\def\eslash{{\hbox{$E$\kern-0.6em\lower-.05ex\hbox{/}\kern0.10em}}}  
\def\vecmet{\mbox{$\vec{\eslash}_T$}} 
\def\vecet{\mbox{$\vec{E}_\text{T}$}} 
\def\MET{\mbox{$\eslash_\text{T}$}\xspace}
\def\met{\mbox{$\eslash_\text{T}$}\xspace}
\def\mex{\mbox{$\eslash_\text{x}$}} 
\def\mey{\mbox{$\eslash_\text{y}$}} 
\def\mepar{\mbox{$\eslash_\parallel$}}
\def\meperp{\mbox{$\eslash_\perp$}}
\def\Zmm{Z \rightarrow \mu\mu}
\def\metvec{\mbox{$\vec{\met}$}\xspace}
\def\metvecrec{\mbox{$\vec{\met}^{\rm rec}$}\xspace}
\def\metvecgen{\mbox{$\vec{\met}^{\rm gen}$}\xspace}
\def\metgen{\mbox{$\met^{\rm gen}$}\xspace}
\def\metparl{\mbox{$\mepar^{\rm rec}$}\xspace}
\def\metperp{\mbox{$\meperp^{\rm rec}$}\xspace}
\def\deltamet{\mbox{$\Delta\met$}\xspace}
\def\pthat{\mbox{$\hat{p}_T$}\xspace}
\def\hslash{{\hbox{$H$\kern-0.8em\lower-.05ex\hbox{/}\kern0.10em}}}
\def\MHT{\mbox{$\hslash_\text{T}$}\xspace}
\def\mht{\mbox{$\hslash_\text{T}$}\xspace}
\def\sumet{\mbox{$\sum \rm{E}_\text{T}$}\xspace}
\def\scalht{\mbox{$H_\text{T}$}\xspace}
\def\etmiss{\mbox{$\eslash_\text{T}$}\xspace}
\def\htmiss{\mbox{$\hslash_\text{T}$}\xspace}
\def\mtt{\mbox{$\rm{M}_\text{T2}$}\xspace}
\def\rmec{\mbox{$R_{\mht/\met}$}\xspace}
\def\bdphi{\mbox{$\Delta\phi^{*}$}\xspace}
\def\bigeslash{{\hbox{$E$\kern-0.38em\lower-.05ex\hbox{/}\kern0.10em}}}  
\def\bigmet{\mbox{$\bigeslash_T$}}
\def\bighslash{{\hbox{$H$\kern-0.6em\lower-.05ex\hbox{/}\kern0.10em}}}  
\def\bigmht{\mbox{$\bighslash_T$}} 
\def\incl{\includegraphics[width=0.49\linewidth]}
\def\inclrot{\includegraphics[angle=90,width=0.47\linewidth]}
\def\INCL{\includegraphics[angle=90,width=0.45\linewidth]}
\def\Incl{\includegraphics[angle=90,width=0.60\linewidth]}

\ifthenelse{\boolean{cms@external}}{\newcommand{\xright}{bottom}}{\newcommand{\xright}{right}}
\ifthenelse{\boolean{cms@external}}{\newcommand{\xleft}{top}}{\newcommand{\xleft}{left}}

\cmsNoteHeader{EWK-10-014} 
\title{\texorpdfstring{Measurement of the Polarization of \PW~Bosons with Large Transverse Momenta in \PW+Jets Events at the LHC}{Measurement of the Polarization of W Bosons with Large Transverse Momenta in W+Jets Events at the LHC}}

\address[cern]{CERN}
\author[cern]{The CMS Collaboration}

\date{\today}

\abstract{
 A first measurement of the polarization of \PW~bosons with large transverse momenta in \Pp\Pp~collisions is presented. The measurement is based on 36~pb$^{-1}$ of data recorded at $\sqrt{s} =$~7~TeV by the CMS detector at the LHC. The left-handed, right-handed and longitudinal polarization fractions ($f_L$, $f_R$, $f_0$) of \PW~bosons with transverse momenta larger than 50~GeV are determined using decays to both electrons and muons. The muon final state yields the most precise measurement, $(f_{L} - f_{R})^{-} = 0.240\,\pm\,0.036\,\textrm{(stat.)}\,\pm\,0.031\,\textrm{(syst.)}$ and $f_{0}^{-} = 0.183\,\pm\,0.087\,\textrm{(stat.)}\,\pm\,0.123\,\textrm{(syst.)}$ for negatively charged \PW~bosons, and $(f_{L} - f_{R})^{+} = 0.310\,\pm\,0.036\,\textrm{(stat.)}\,\pm\,0.017\,\textrm{(syst.)}$ and $f_{0}^{+} = 0.171\,\pm\,0.085\,\textrm{(stat.)}\,\pm\,0.099\,\textrm{(syst.)}$ for positively charged \PW~bosons. This establishes, for the first time, that \PW~bosons produced in \Pp\Pp~collisions with large transverse momenta are predominantly left-handed, as expected in the standard model.
}

\hypersetup{%
pdfauthor={CMS Collaboration},%
pdftitle={Measurement of the Polarization of W Bosons with Large Transverse Momenta in W+Jets Events at the LHC},%
pdfsubject={CMS},%
pdfkeywords={CMS, physics}}

\maketitle 

The measurement of the kinematic properties of \PW~bosons produced at hadron colliders provides a stringent test of perturbative quantum chromodynamics (QCD) calculations as well as being an important prerequisite to searches for physics beyond the standard model.  The \Pp\Pp~collisions at the Large Hadron Collider (LHC) offer both a new environment and higher energy to study \PW~bosons with large transverse momenta recoiling against several energetic jets. The sizable production cross section results in significant samples of \PW~bosons, while the nature of the initial state leads to an enhancement of the quark-gluon contribution to \PW+jet production when compared to the Tevatron $\Pp\Pap$~collider, where quark-gluon and antiquark-gluon processes contribute equally. This dominance of quark-gluon initial states, along with the $V-A$ nature of the coupling of the \PW~boson to fermions, implies that at the LHC \PW~bosons with high transverse momenta are expected to exhibit a sizable left-handed polarization. A significant asymmetry in the transverse momentum spectra of the neutrino and charged lepton from subsequent leptonic \PW~ decays is therefore expected. This Letter reports the first measurement of the polarization of \PW~bosons with large transverse momenta at the LHC, using a data sample of \Pp\Pp~collisions corresponding to an integrated luminosity of $36 \pm 1.4$~pb$^{-1}$ at a center-of-mass energy of 7 TeV, recorded with the Compact Muon Solenoid (CMS) detector.

We measure the polarization of the \PW~boson in the helicity frame, where the polar angle ($\theta^{*}$) of the charged lepton from the decay in the \PW~rest frame is measured with respect to the boson flight direction in the laboratory frame. The azimuthal angle ($\phi^{*}$) is defined to be zero for the proton which has the smaller $\theta^{*}$ in the boson rest frame. The cross section for \PW~production at a hadron collider with a subsequent leptonic decay, $\textrm{d}N/\textrm{d}\Omega$, is given by~\cite{mirkes}

\begin{equation}
\label{diffxsec}
\begin{split}
\frac{\textrm{d}N}{\textrm{d}\Omega} \propto (1+\cos^2 \theta^*) + \tfrac{1}{2}A_0(1-3\cos^2\theta^*) + A_1\sin2\theta^*\cos\phi^* \\
+\,\tfrac{1}{2}A_2\sin^2\theta^*\cos2\phi^* + A_3\sin\theta^*\cos\phi^* + A_4\cos\theta^*,
\end{split}
\end{equation}
where the coefficients $A_{i}$~$(i=0,\ldots,4)$ depend on the \PW~boson charge, transverse momentum and rapidity, and make up the elements of the polarization density matrix. Integrating Eq. (\ref{diffxsec}) over $\phi^{*}$ yields

\begin{equation}
\label{hframe_xsecminus}
\frac{\textrm{d}N}{\textrm{d}\cos\theta^*} \propto (1+\cos^{2} \theta^*) + \tfrac{1}{2}A_{0}(1-3\cos^{2}\theta^*) + A_{4}\cos\theta^*.
\end{equation}

The fractions of left-handed, right-handed, and longitudinal \PW~bosons ($f_{L}, f_{R}$ and $f_{0}$, respectively) are related to the parameters $A_{i}$ by $A_{0} \propto f_{0}$ and $A_{4} \propto \pm(f_{L} - f_{R})$ depending on the \PW~boson charge, where by definition $f_i>0$ and $f_L + f_R + f_0 = 1$. {\em A priori}, the values of the $f_i$ parameters are not expected to be the same for both charges, since for partons which carry a large fraction of the proton's momentum, the ratio of valence $u$~quarks to sea quarks is higher than that for valence $d$~quarks.

The amount of \PW~boson momentum imparted to the charged decay lepton is determined by $\cos\theta^*$, and hence an asymmetry in the $\cos\theta^{*}$ distribution leads to an asymmetry between the neutrino and charged-lepton momentum spectra. This can be quantified via a measurement of the $A_4$ parameter. However, the inability to determine the momentum of the neutrino along the beam axis introduces a two-fold ambiguity in the determination of the momentum of the \PW~boson. Therefore, it is not possible to precisely determine the \PW~boson rest frame required to extract the \PW~decay angles. To overcome this, a variable which exhibits a strong correlation with $\cos\theta^*$ is introduced. The lepton projection variable, $L_P$, is defined as the projection of the scaled transverse momentum of the charged lepton, $\vec p_{T}(\ell)/|\vec p_{T}(\PW)|$, onto the normalized transverse momentum of the parent \PW~boson, $\vec p_{T}(\PW)/|\vec p_{T}(\PW)|$:

\begin{equation}
\label{LPdefinition}
L_{P} = \frac{\vec p_{T}(\ell) \cdot \vec p_{T}(\PW)}{|\vec p_{T}(\PW)|^2}.
\end{equation}

In the above expression, $\vec p_{T}(\PW)$ is estimated from the vectorial sum of the missing transverse energy \vecmet~and $\vec p_{T}(\ell)$ in the event. Experimentally, \vecmet~is reconstructed as the negative vector sum of the transverse energy vectors of all particles identified in the event using a particle flow algorithm~\cite{PFMET}. In the limit of very high $p_T(\PW)$, $L_P$ lies within the range [0,1] and $\cos\theta^* = 2(L_P - \frac{1}{2})$.

The central feature of the CMS apparatus is a superconducting solenoid, 13~m in length and 6~m in diameter, which provides an axial magnetic field of 3.8~T. The bore of the solenoid is instrumented with various particle detection systems. Charged particle trajectories are measured by the silicon pixel and strip tracking detectors, covering $0 < \phi < 2\pi$ in azimuth and $|\eta| <$ 2.5, where the pseudorapidity is defined as $\eta = -\ln[\tan (\theta/2)]$, and $\theta$ is the polar angle of the trajectory of the particle with respect to the counterclockwise beam direction. A crystal electromagnetic calorimeter (ECAL) and a brass/scintillator hadron calorimeter (HCAL) surround the tracking volume and cover the region $|\eta| < 3$. The steel return yoke outside the solenoid is in turn instrumented with gas detectors which are used to identify muons. The detector is nearly hermetic, allowing for energy balance measurements in the plane transverse to the beam direction. A more detailed description of the CMS detector can be found elsewhere~\cite{ref:CMS}.

The trigger providing the data sample used in this analysis is based on the presence of at least one charged lepton, either an electron or a muon, with a minimum transverse momentum of 22~(15)~GeV for the electron (muon). Events passing this trigger are required to have at least one good reconstructed \Pp\Pp~interaction vertex~\cite{CMS-PAS-TRK-10-001}.  Electrons and muons are reconstructed and selected using the procedure and requirements described in the measurement of the inclusive $\PW/\cPZ$~boson cross section~\cite{WZ-xsec-paper}.  The selection of \PW~boson candidates requires one electron (muon), with $p_T > 25\,(20)$~GeV in $|\eta|<2.4\,(2.1)$. High-$p_{T}$ leptons are also found in events in which hadronic jets mimic the lepton signature. Such misidentified leptons, as well as non prompt leptons arising from decays of heavy-flavor hadrons or decays of light mesons within jets, are suppressed by imposing limits on the additional hadronic activity surrounding the lepton candidate in an event. The scalar sum of the transverse momenta of all charged particle tracks and the transverse energy in the ECAL and HCAL in a cone of $\Delta R=\sqrt{(\Delta\phi)^2 + (\Delta\eta)^2} = 0.3$ centered on the lepton candidate is calculated, excluding the contribution from the candidate itself. The candidate is retained if this sum is less than 4~(10)\% of the electron (muon) $p_T$. Electrons (muons) from decays of \cPZ~bosons are suppressed by vetoing events containing a second lepton with $p_T > 15\,(10)$~GeV passing looser isolation criteria.

Since the analysis measures the lepton and neutrino momenta from \PW~boson decays, there is no requirement on the~\vecmet~in the event. Instead, to further reduce backgrounds from QCD multijet production, the selection requires $M_T > 50\,(30)$~GeV for the electron (muon) channel, where $M_T = \sqrt{2|\vec{p}_{T}(\ell)||\vecmet|(1-\cos\Delta\phi)}$ and $\Delta\phi$ is the angle between the missing transverse momentum and the lepton transverse momentum. The requirement on $M_T$ is higher in the electron channel to compensate for the larger QCD multijet background. Given that the polarization and correlation of $L_P$ with $\cos\theta^{*}$ increase with $p_T (\PW)$, while the number of available events decreases sharply with $p_T (\PW)$, we require $p_T (\PW) > 50$~GeV as the result of an optimization study based on the expected statistical uncertainty of the $(f_{L} - f_{R})$ measurement. As high-$p_T$ \PW~bosons are also produced in top quark decays, only events with up to three reconstructed jets are retained. The jets considered are particle-flow based~\cite{ref:pfjets} with $p_{T}>30$~GeV, $|\eta| < 5$, and are clustered using the anti-k$_\text{T}$ algorithm~\cite{ref:antikt} with a distance parameter of 0.5.
In data, a total of 5485 (8626) events pass the selection requirements in the electron (muon) channel. These events are almost entirely \PW+jets events, with a small contamination from the processes \ttbar+jets, \cPZ+jets and photon+jets. All these processes, and their expectations, are produced using the \MADGRAPH~\cite{madgraph, pythia} generator, with the {\sc CTEQ6L}~\cite{cteq} parton distribution function set, and are passed through a full simulation of the CMS detector based on the \GEANTfour~\cite{ref:geant4} package. There are 252~$\pm$~93~(266~$\pm$~84) estimated background events from simulation in the electron (muon) channel, where the uncertainty corresponds to the theoretical uncertainty on the relevant cross sections.

In the muon channel, the background from QCD multijet and heavy flavor production is expected to be negligible. In the electron channel, the simulation predicts a higher level of multijet background, and therefore the distribution of the $L_P$ variable for the surviving background events is needed. This distribution is obtained using data enriched in misidentified electrons by reversing some of the electron selection requirements, as in~\cite{WZ-xsec-paper}. We refer to this as the ``antiselected sample''. As a cross-check, the procedure is also applied to simulated samples. The $L_P$ distribution from the QCD multijet background after all selection cuts is found to be well reproduced by the antiselected electron sample.

The polarization fraction parameters~$(f_{L}-f_R)$~and~$f_{0}$~are measured using a binned maximum likelihood fit to the~$L_P$~variable, separately for~$\PW^{+}$~and~$\PW^{-}$~bosons in the electron and muon final states. The $L_P$ distribution for each of the three polarization states of the \PW~boson is extracted from Monte Carlo samples which are reweighted to the angular distributions expected from each polarization state in the \PW~boson center-of-mass frame. The $L_P$ distributions are simulated in the presence of pile-up events matching the vertex multiplicity distribution observed in data, corresponding to an average of $2.8$~reconstructed vertices per event.

\begin{figure}
\begin{center}
  \includegraphics[width=0.49\textwidth,angle=0]{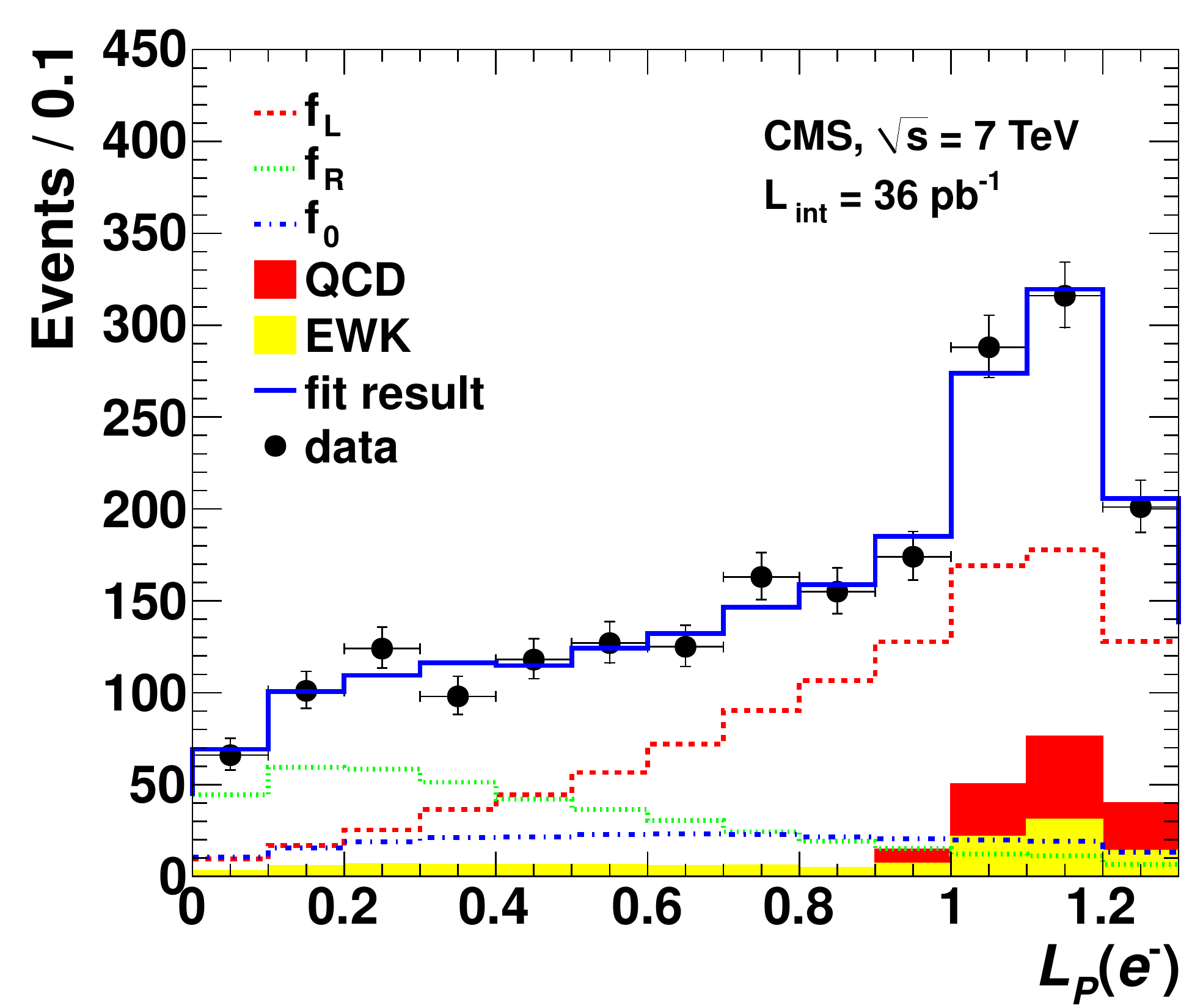}
  \includegraphics[width=0.49\textwidth,angle=0]{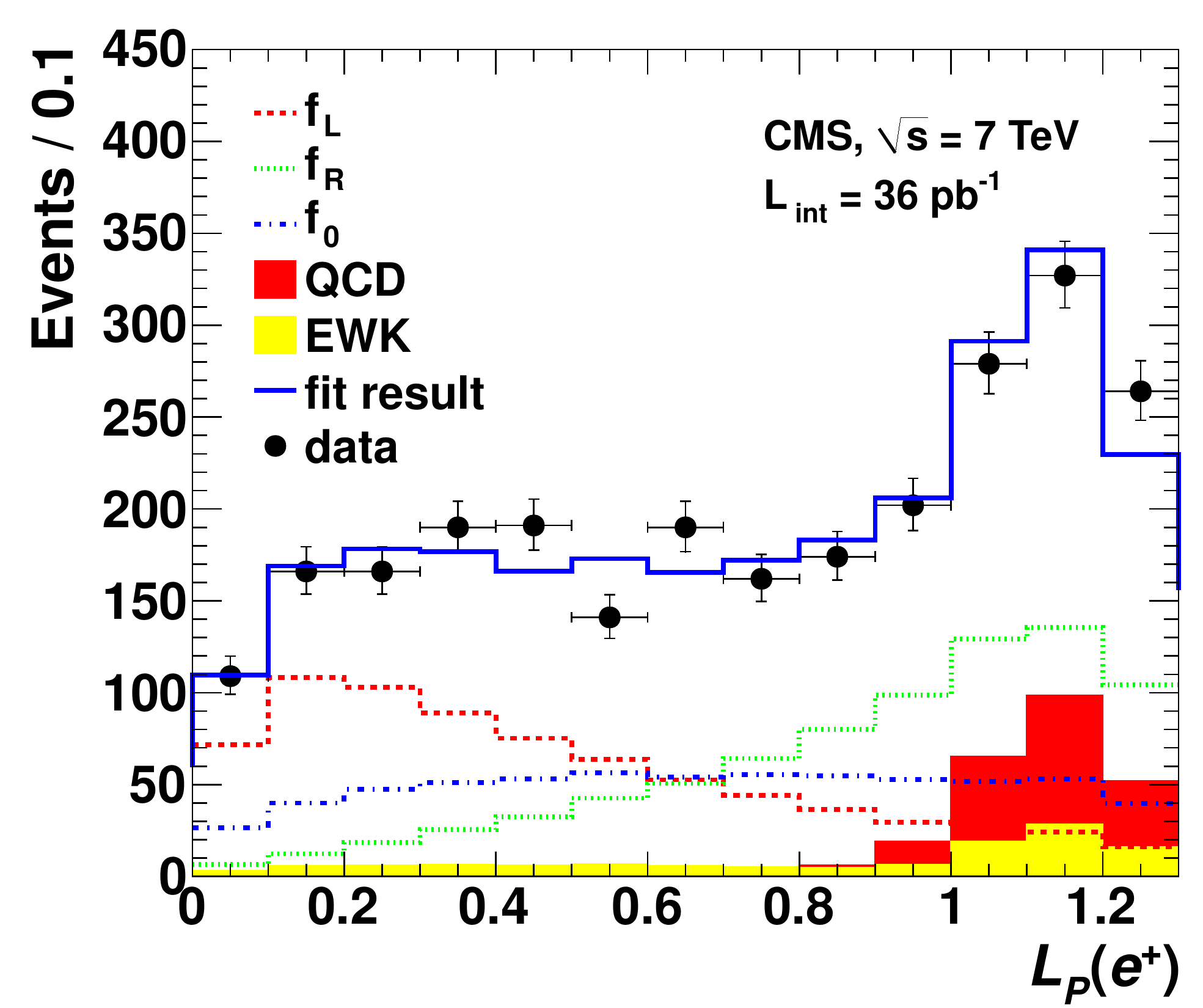}
       \caption{Fit results using 36~$\textrm{pb}^{-1}$ of collision data for the $L_P(\Pem)$ (\xleft) and $L_P(\Pep)$ (\xright) distributions. The left-handed, right-handed and longitudinal \PW~components, with normalization as determined by the fit, are represented by the dashed, dotted, and dash-dotted lines respectively. The shaded distributions show the QCD and EWK backgrounds. The solid line represents the sum of all individual components, and can be directly compared with the data distribution (circles). }
   \label{fig:eLPdatafit}
\end{center}
\end{figure}

\begin{figure}
\begin{center}
  \includegraphics[width=0.49\textwidth,angle=0]{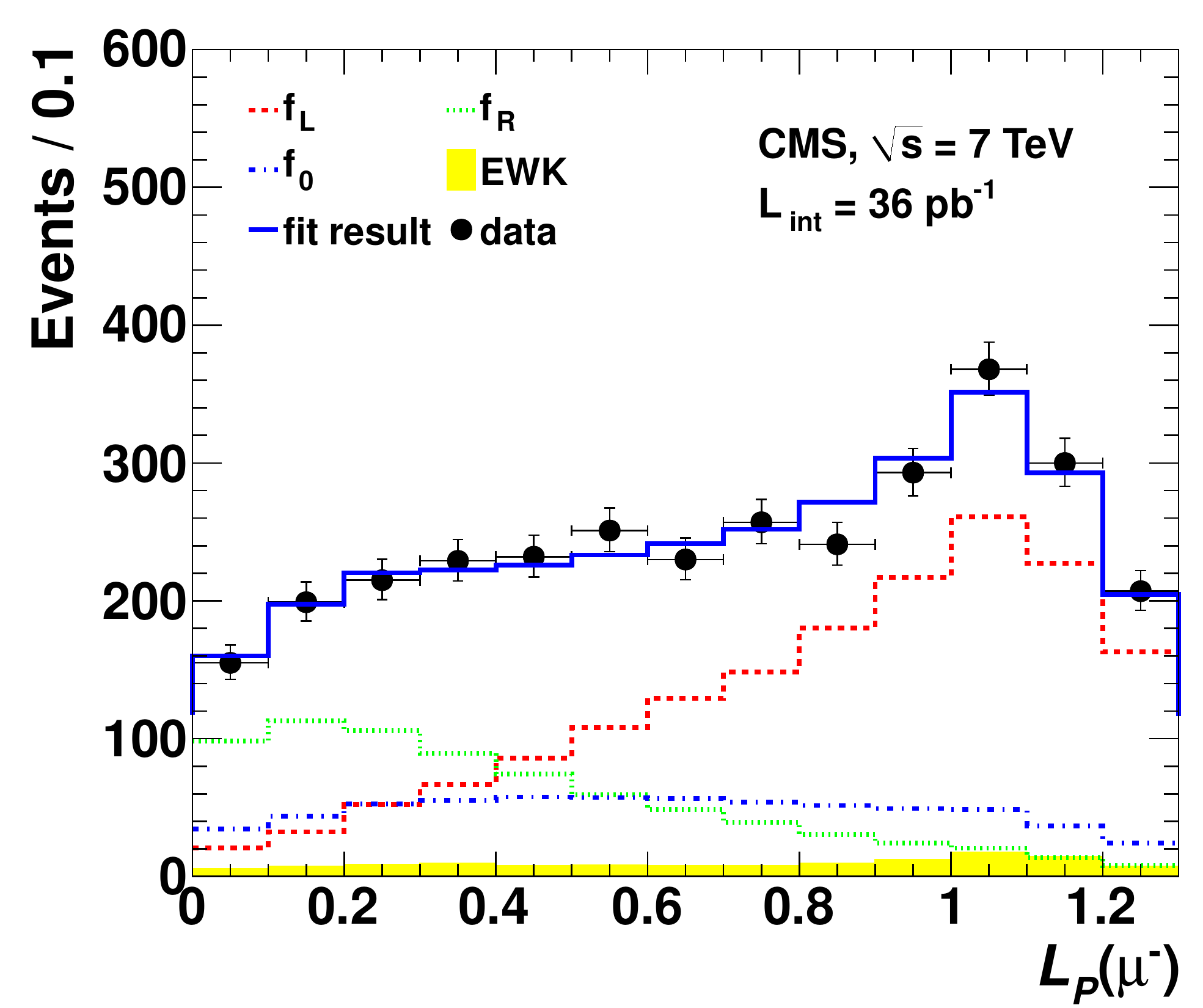}
  \includegraphics[width=0.49\textwidth,angle=0]{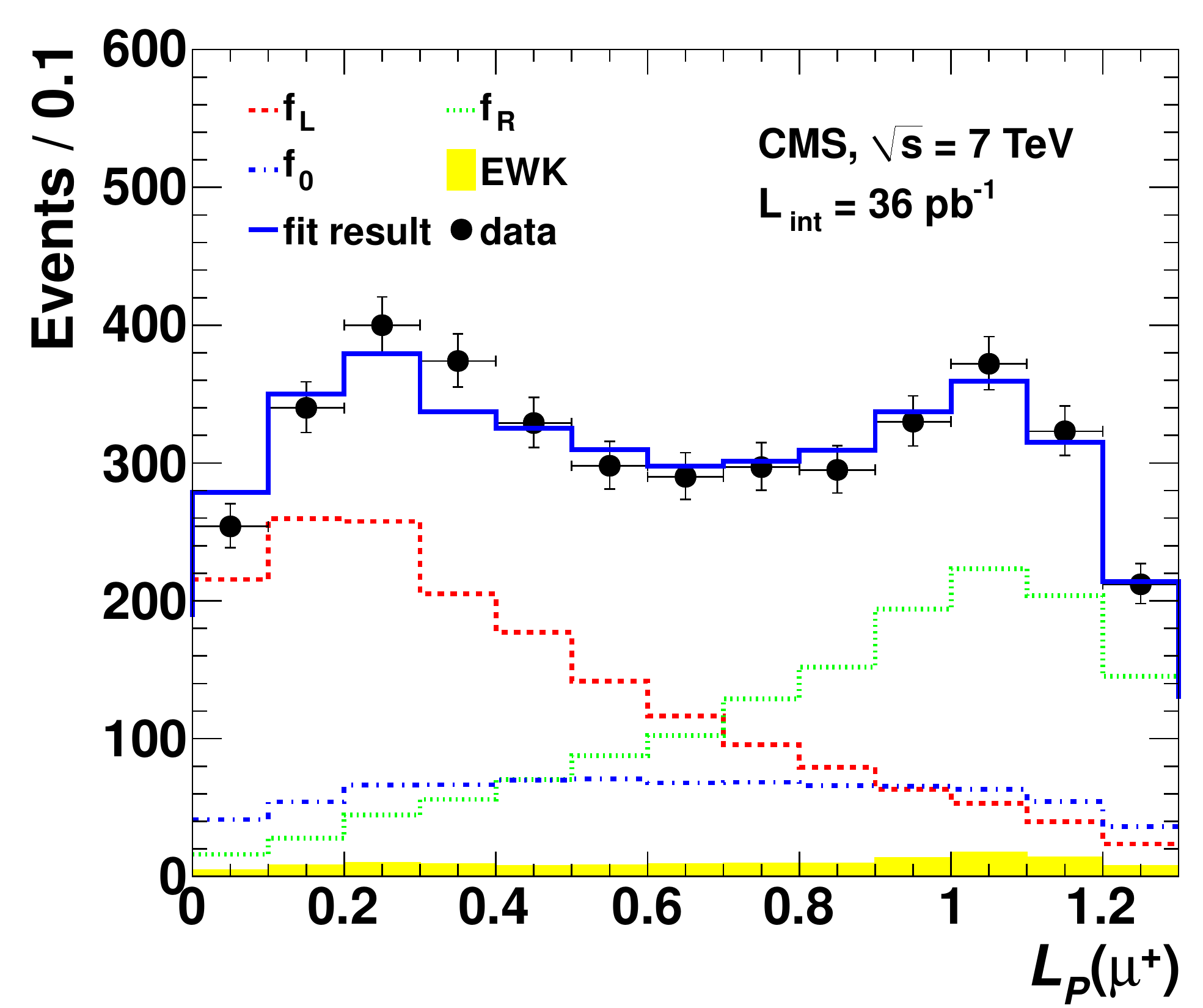}
       \caption{Fit results using 36~$\textrm{pb}^{-1}$ of collision data for the $L_P(\Pgmm)$ (\xleft) and $L_P(\Pgmp)$ (\xright) distributions. The left-handed, right-handed and longitudinal \PW~components, with normalization as determined by the fit, are represented by the dashed, dotted, and dash-dotted lines respectively. The shaded distribution shows the EWK backgrounds. The solid line represents the sum of all individual components, and can be directly compared with the data distribution (circles). }
   \label{fig:mLPdatafit}
\end{center}
\end{figure}

The $L_P$ distributions for electrons and muons are shown in Figs.~\ref{fig:eLPdatafit} and~\ref{fig:mLPdatafit}, respectively. Also shown are the results of the fit to the individual components corresponding to the three \PW~polarization states, and to the background. The background consists of an electroweak component and a QCD multijet component, which is negligible in the muon sample. The fit is carried out by keeping the electroweak background contribution fixed to the value predicted by simulation, whereas all other components, including the QCD multijet background, are allowed to vary. The results of the fits, along with the correlations between these extracted parameters, are listed for positively and negatively charged electrons and muons in Table~\ref{tab:datafitsummary}. For each \PW~boson charge, the results for electrons and muons are self-consistent. The correlations differ due to the QCD multijet component included in the fit to the electron final state. Also shown are the results from performing a combined fit, simultaneously to both the electron and muon data.

\begin{table}
  \caption{A summary of the individual and combined fit results for negatively charged and positively charged electrons and muons. Systematic uncertainties from different sources are combined in quadrature.}\label{tab:datafitsummary}
  \begin{center}
{\footnotesize
    \begin{tabular}{| l | c | c | c |} \hline
                                  &  Fit result                                                              \\\hline
 $\Pe: (f_{L} - f_{R})^{-}$       & $0.187 \pm 0.069$ (stat.)  $\pm 0.066$ (syst.)                           \\\hline
 $\Pe: f_{0}^{-}$                 & $0.130 \pm 0.200$ (stat.)  $\pm 0.174$ (syst.)                           \\\hline
 Correlation                      & $-0.204$ (stat.), $-0.283$ (stat. + syst.)                               \\\hline\hline
 $\Pe: (f_{L}-f_{R})^{+}$         & $0.277 \pm 0.060$ (stat.)  $\pm 0.050$ (syst.)                           \\\hline
 $\Pe: f_{0}^{+}$                 & $0.240 \pm 0.190$ (stat.)  $\pm 0.090$ (syst.)                           \\\hline
 Correlation                      & $-0.295$ (stat.), $0.001$ (stat. + syst.)                                \\\hline\hline
 $\Pgm: (f_{L} - f_{R})^{-}$      & $0.240 \pm 0.036$ (stat.) $\pm 0.031$ (syst.)                            \\\hline
 $\Pgm: f_{0}^{-}$                & $0.183 \pm 0.087$ (stat.) $\pm 0.123$ (syst.)                            \\\hline
 Correlation                      & $0.395$ (stat.), $-0.308$ (stat. + syst.)                                \\\hline\hline
 $\Pgm: (f_{L}-f_{R})^{+}$        & $0.310 \pm 0.036$ (stat.) $\pm 0.017$ (syst.)                            \\\hline
 $\Pgm: f_{0}^{+}$                & $0.171 \pm 0.085$ (stat.) $\pm 0.099$ (syst.)                            \\\hline
 Correlation                      & $-0.721$ (stat.), $-0.269$ (stat. + syst.)                               \\\hline\hline
 Combined: $(f_{L} - f_{R})^{-}$  & $0.226 \pm 0.031$ (stat.) $\pm 0.050$ (syst.)                            \\\hline
 Combined: $f_{0}^{-}$            & $0.162 \pm 0.078$ (stat.) $\pm 0.136$ (syst.)                            \\\hline
 Correlation                      & $0.304$ (stat.), $-0.326$ (stat. + syst.)                                \\\hline\hline
 Combined: $(f_{L}-f_{R})^{+}$    & $0.300 \pm 0.031$ (stat.) $\pm 0.034$ (syst.)                            \\\hline
 Combined: $f_{0}^{+}$            & $0.192 \pm 0.075$ (stat.) $\pm 0.089$ (syst.)                            \\\hline
 Correlation                      & $-0.660$ (stat.), $-0.121$ (stat. + syst.)                               \\\hline

    \end{tabular}
}
  \end{center}
\end{table}

Several experimental and theoretical effects are considered as sources of systematic uncertainty. The most significant sources, which are listed in Table~\ref{tab:systematic}, stem from the recoil energy scale and resolution~\cite{PAS-JME-10-005} uncertainties, which enter in the measurement of the transverse momentum of the \PW~boson. The recoil energy scale is varied by its measured uncertainty~\cite{JME-10-010} and the effect is propagated through the analysis, resulting in modified $L_P$ distributions. The measurement is repeated and the full difference from the nominal value is quoted as the systematic uncertainty from this source. The effect is smaller for values of $L_P$ close to one, corresponding to low values of~\vecmet, and hence the uncertainty is smaller for $\PW^{-}$~relative to $\PW^{+}$. The same procedure is followed for the recoil resolution, electron energy, and muon momentum scale. Decays of \cPZ~bosons to electrons are used to derive corrections, in bins of the electron pseudorapidity, which calibrate the electron energy scale. An uncertainty of~$\pm\,50$\% on these corrections is assumed, in order to cover the full range of variations. Decays of \cPZ~bosons to muons are used to constrain the muon momentum scale and an uncertainty of 1\% at 100~GeV is found. The fit range of the lepton projection variable is restricted to $0.0 < L_P < 1.3$, as a result of the minimization of the combined statistical and systematic uncertainties of the measurement.

The uncertainty on the modeling of the QCD background in the electron channel is estimated using the sample of antiselected electrons which yields the shape of the $L_P$ distribution for this background. The fit is repeated multiple times, whilst varying the $L_P$ distribution of the antiselected sample within its statistical uncertainties. The variation in the fit results is then used as an estimate of the systematic uncertainty, which is found to be negligible when compared to the leading systematic uncertainties.

A mismeasurement of the lepton charge dilutes the measurement of the \PW~boson polarization. The misidentification rate is studied as a function of pseudorapidity using \cPZ~bosons decaying into a pair of oppositely charged leptons. This effect is found to be negligible for both electron and muon channels.

The systematic uncertainty arising from matching the vertex multiplicity distribution in the simulation to that observed in the data is estimated by varying the former within the statistical uncertainty of the latter, and is found to be negligible.

The effect of the theoretical uncertainties on the normalization of the electroweak background distributions, corresponding to $25\%$ for the \cPZ~boson and $50\%$ for the top quark, is included in the fit and found to contribute a negligible systematic uncertainty to the \PW~boson polarization measurement. The lepton projection variable also depends weakly on the values of the polarization parameters $A_1, A_2$ and $A_3$, which are not measured. In order to evaluate the magnitude of the effect, these coefficients are varied by~$\pm\,10$\% with respect to recent standard model calculations at leading-order QCD~\cite{ref:lancepaper}. These variations produce a negligible change in the \PW~boson polarization measurement. A similar result is obtained for the shape of the $L_P$ distributions by varying the parton distribution functions using the~{\sc CTEQ6.6}~PDF error set.

\begin{table}
  \caption{Summary of the leading systematic uncertainties for the electron and muon final states, as well as for the combined measurement. The total systematic uncertainties are also shown for reference.}\label{tab:systematic}
  \begin{center}
{\footnotesize
    \begin{tabular}{| c | c | c | c |c |} \hline
Uncertainty & $(f_{L} - f_{R})^{-}$  & $f_{0}^{-}$  & $(f_{L} - f_{R})^{+}$  & $f_{0}^{+}$ \\  \hline \hline
& \multicolumn{4}{|c|}{Electron channel}  \\ \hline
Recoil energy scale & $\pm 0.042$    &  $\pm 0.150$ & $\pm 0.027$            & $\pm 0.078$ \\ \hline
Recoil resolution   & $\pm 0.046$    &  $\pm 0.047$ & $\pm 0.037$            & $\pm 0.039$ \\ \hline
Electron scale      & $\pm 0.017$    &  $\pm 0.014$ & $\pm 0.019$            & $\pm 0.016$ \\ \hline
Total uncertainty   & $\pm 0.066$    &  $\pm 0.174$ & $\pm 0.050$            & $\pm 0.090$ \\ \hline \hline
& \multicolumn{4}{|c|}{Muon channel}  \\ \hline
Recoil energy scale & $\pm 0.029$    &  $\pm 0.123$ & $\pm 0.011$            & $\pm 0.092$ \\ \hline
Recoil resolution   & $\pm 0.012$    &  $\pm 0.006$ & $\pm 0.012$            & $\pm 0.004$ \\ \hline
Muon scale          & $\pm 0.002$    &  $\pm 0.007$ & $\pm 0.004$            & $\pm 0.008$ \\ \hline
Total uncertainty   & $\pm 0.031$    &  $\pm 0.123$ & $\pm 0.017$            & $\pm 0.099$ \\ \hline \hline
& \multicolumn{4}{|c|}{Combined measurement}  \\ \hline
Recoil energy scale & $\pm 0.033$    &  $\pm 0.133$ & $\pm 0.016$            & $\pm 0.087$ \\ \hline
Recoil resolution   & $\pm 0.035$    &  $\pm 0.023$ & $\pm 0.027$            & $\pm 0.015$ \\ \hline
Electron scale      & $\pm 0.013$    &  $\pm 0.011$ & $\pm 0.012$            & $\pm 0.008$ \\ \hline
Muon scale          & $\pm 0.002$    &  $\pm 0.004$ & $\pm 0.004$            & $\pm 0.004$ \\ \hline
Total uncertainty   & $\pm 0.050$    &  $\pm 0.136$ & $\pm 0.034$            & $\pm 0.089$ \\ \hline
     \end{tabular}
}
  \end{center}
\end{table}

\begin{figure}
\begin{center}
  \includegraphics[width=0.49\textwidth,angle=0]{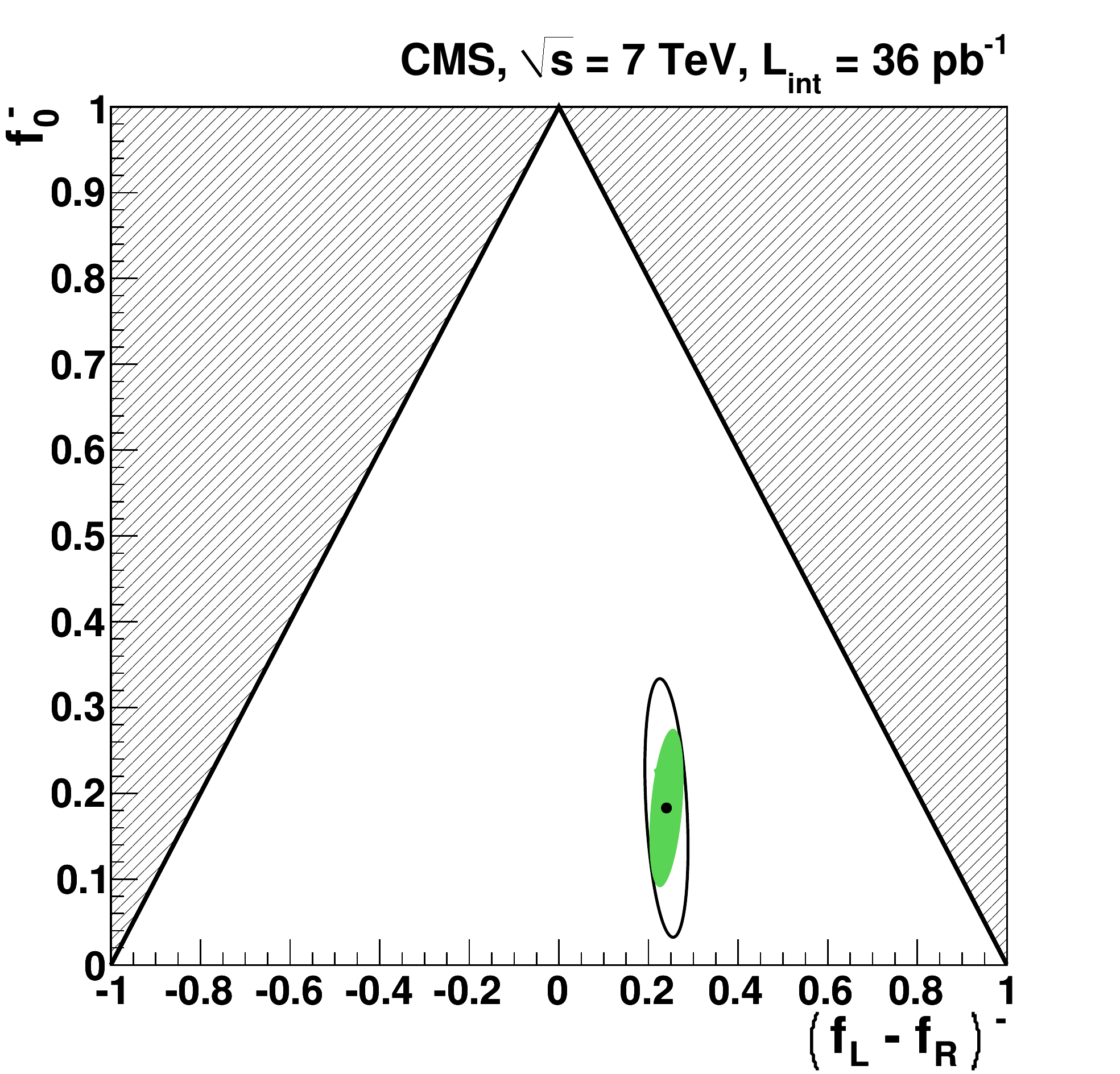}
  \includegraphics[width=0.49\textwidth,angle=0]{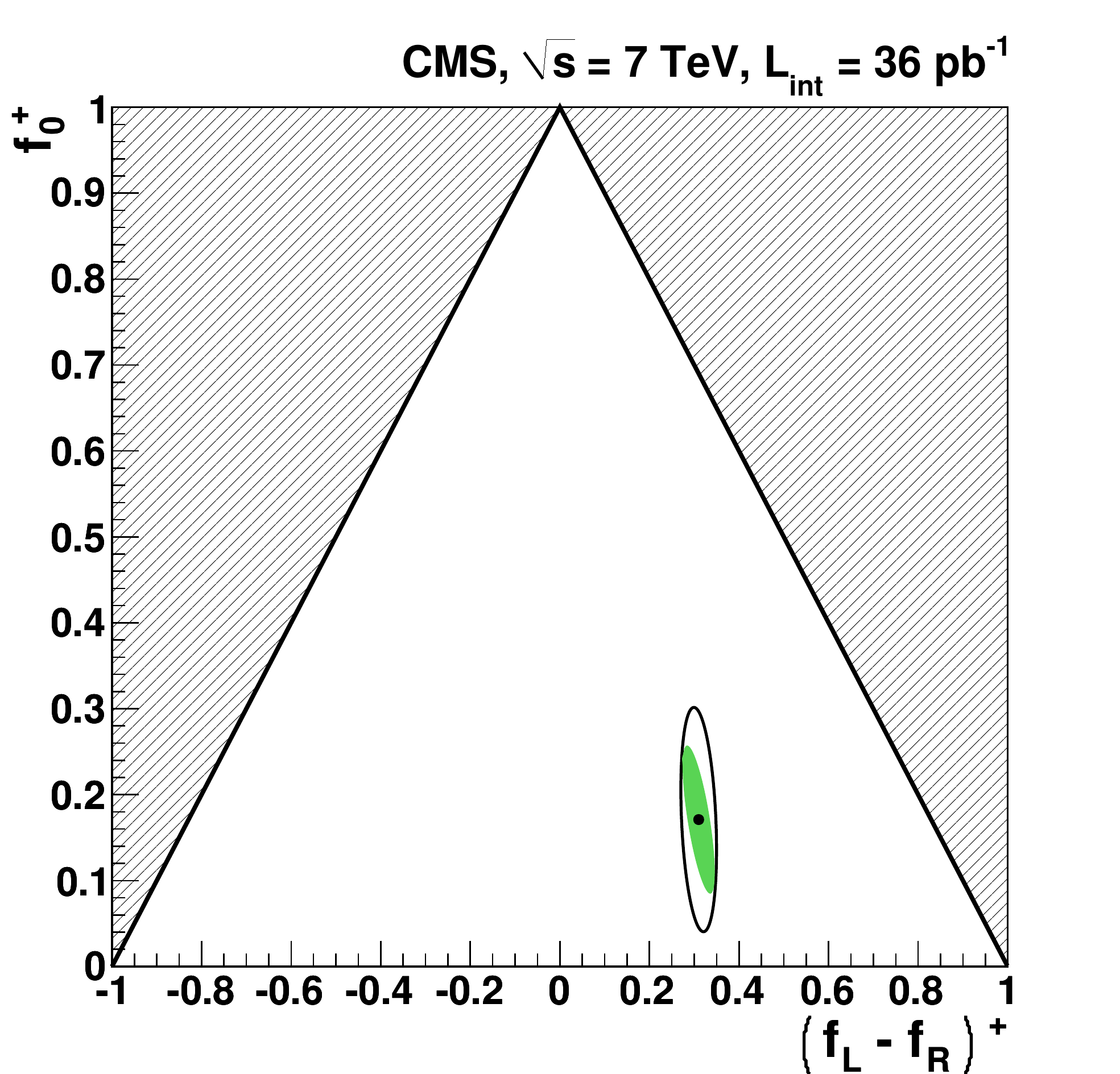}
       \caption{The muon fit result (black dot) in the $((f_{L} - f_{R}), f_0)$ plane for negatively charged (\xleft) and positively charged (\xright) leptons. The 68\% confidence level contours for the statistical and total uncertainties are shown by the green shaded region and the black contour respectively. The disallowed region is hatched. }
   \label{fig:combContours}
\end{center}
\end{figure}

The muon fit result, having the smallest total uncertainty, is shown in the $((f_{L} - f_{R}), f_0)$ plane for each \PW~charge in Fig.~\ref{fig:combContours}. The 68\% confidence level contours for both the statistical and total uncertainties are also shown. With the current sensitivity, the values of $(f_{L} - f_{R})$ and $f_0$ do not differ significantly for $\PW^+$~and $\PW^-$. When compared to recent standard model calculations~\cite{ref:lancepaper}, the results agree well.

In conclusion, the first measurement of the polarization of \PW~bosons with large transverse momenta at a \Pp\Pp~collider has been presented. Using a sample of collision data corresponding to an integrated luminosity of 36~pb$^{-1}$, the measurement is performed for both charges of the \PW~boson, in the electron and muon final states. The results from both of these channels are consistent, as are the combined fit results. The muon fit result yields the most precise measurement, $(f_{L} - f_{R})^{-} = 0.240\,\pm\,0.036\,\textrm{(stat.)}\,\pm\,0.031\,\textrm{(syst.)}$ and $f_{0}^{-} = 0.183\,\pm\,0.087\,\textrm{(stat.)}\,\pm\,0.123\,\textrm{(syst.)}$ for negatively charged \PW~bosons, and $(f_{L} - f_{R})^{+} = 0.310\,\pm\,0.036\,\textrm{(stat.)}\,\pm\,0.017\,\textrm{(syst.)}$ and $f_{0}^{+} = 0.171\,\pm\,0.085\,\textrm{(stat.)}\,\pm\,0.099\,\textrm{(syst.)}$ for positively charged \PW~bosons. This measurement establishes a difference between the left-handed and right-handed polarization parameters with a significance of 7.8 standard deviations for $\PW^+$~bosons and 5.1 standard deviations for $\PW^-$~bosons. This is the first observation that high-$p_T$~\PW~bosons produced in \Pp\Pp~collisions are predominantly left-handed, as expected in the standard model.

We wish to congratulate our colleagues in the CERN accelerator departments for the excellent performance of the LHC machine. We thank the technical and administrative staff at CERN and other CMS institutes, and acknowledge support from: FMSR (Austria); FNRS and FWO (Belgium); CNPq, CAPES, FAPERJ, and FAPESP (Brazil); MES (Bulgaria); CERN; CAS, MoST, and NSFC (China); COLCIENCIAS (Colombia); MSES (Croatia); RPF (Cyprus); Academy of Sciences and NICPB (Estonia); Academy of Finland, ME, and HIP (Finland); CEA and CNRS/IN2P3 (France); BMBF, DFG, and HGF (Germany); GSRT (Greece); OTKA and NKTH (Hungary); DAE and DST (India); IPM (Iran); SFI (Ireland); INFN (Italy); NRF and WCU (Korea); LAS (Lithuania); CINVESTAV, CONACYT, SEP, and UASLP-FAI (Mexico); PAEC (Pakistan); SCSR (Poland); FCT (Portugal); JINR (Armenia, Belarus, Georgia, Ukraine, Uzbekistan); MST and MAE (Russia); MSTD (Serbia); MICINN and CPAN (Spain); Swiss Funding Agencies (Switzerland); NSC (Taipei); TUBITAK and TAEK (Turkey); STFC (United Kingdom); DOE and NSF (USA).

\bibliography{auto_generated}   

\cleardoublepage\appendix\section{The CMS Collaboration \label{app:collab}}\begin{sloppypar}\hyphenpenalty=5000\widowpenalty=500\clubpenalty=5000\textbf{Yerevan Physics Institute,  Yerevan,  Armenia}\\*[0pt]
S.~Chatrchyan, V.~Khachatryan, A.M.~Sirunyan, A.~Tumasyan
\vskip\cmsinstskip
\textbf{Institut f\"{u}r Hochenergiephysik der OeAW,  Wien,  Austria}\\*[0pt]
W.~Adam, T.~Bergauer, M.~Dragicevic, J.~Er\"{o}, C.~Fabjan, M.~Friedl, R.~Fr\"{u}hwirth, V.M.~Ghete, J.~Hammer\cmsAuthorMark{1}, S.~H\"{a}nsel, M.~Hoch, N.~H\"{o}rmann, J.~Hrubec, M.~Jeitler, W.~Kiesenhofer, M.~Krammer, D.~Liko, I.~Mikulec, M.~Pernicka, H.~Rohringer, R.~Sch\"{o}fbeck, J.~Strauss, A.~Taurok, F.~Teischinger, P.~Wagner, W.~Waltenberger, G.~Walzel, E.~Widl, C.-E.~Wulz
% ...
\vskip\cmsinstskip
% ...
1:~~Also at CERN, European Organization for Nuclear Research, Geneva, Switzerland\\
% ...

\end{sloppypar}
\end{document}